\documentclass[%
 reprint,%
 amssymb, amsmath,%
 aip,jap,
groupedaddress,%
]{revtex4-1}

\usepackage{graphicx,amsmath}
\usepackage{subfigure}
\usepackage{longtable}
\usepackage{dcolumn}%
\usepackage{bm}%
\usepackage{docs}%

\begin{document}
\title{Sine gating detector with simple filtering for low-noise infra-red single photon detection at room temperature}
\author{Nino Walenta}
\email{nino.walenta@unige.ch}
\author{Tommaso Lunghi}
\author{Olivier Guinnard}
\author{Raphael Houlmann}
\author{Hugo Zbinden}
\author{Nicolas Gisin}
\affiliation{Group of Applied Physics-Optique, University of Geneva, Chemin de Pinchat 22, 1211 Geneva, Switzerland}
\date{\today}
\begin{abstract}
We present and analyze a gated single photon avalanche detector using a sine gating scheme with a simple but effective low-pass filtering technique for fast low-noise single photon detection at telecom wavelength. The detector is characterized by 130~ps short gates applied with a frequency of 1.25~GHz, yields only 70~ps timing jitter and noise probabilities as low as 7$\cdot$10$^{\text{-7}}$ per gate at 10~\% detection efficiency. We show that the detector is suitable for high rate quantum key distribution (QKD) and even at room temperature it could allow for QKD over distances larger than 25~km.
\end{abstract}
\maketitle
Quantum key distribution (QKD)~\cite{Bennett84,Gisin2002b} is the most complex and advanced application of quantum physics adopted commercially today. Improving the performance and  transmission distances of direct point-to-point quantum communication systems in optical telecom networks is a challenge where near-infrared single photon detectors play a crucial role. Also other applications based on single photons such as optical fiber metrology \cite{Eraerds2010a,Healey1980}, optical remote sensing \cite{Priedhorsky:96,Warburton:07,Ren:11} and distributed quantum computation require reliable, efficient single photon detectors with high detection rates, low timing jitter and low noise contributions due to dark count or afterpulsing. Although in the past years alternative detection techniques, e.g. based on cryogenically cooled superconducting nanowires \cite{goltsman:705,SSPD2007,Stucki2009a} or sum-frequency generation \cite{Thew2006,Thew2009}, have been successfully implemented, InGaAs/InP single photon avalanche diodes (SPADs) are today's preferred choice for detectors in integrated systems due to their compactness, robustness and reliability \cite{Thew2009,Itzler2011}.

Besides their advantages, one of the factors limiting the performance of InGaAs/InP SPADs is the afterpulsing effect and several approaches have been taken to mitigate the impairment in quantum communication setups \cite{COVA91, Kindt99, Itzler2011}. Afterpulses appear due to impurities in the semiconductor, which with a certain probability can trap charge carriers generated during a photon induced avalanche, and release them at a later time depending on the lifetime of the relevant afterpulse trap type. Recently, new gating schemes have been proposed and implemented which are based on very short gating periods during which the detector is sensitive for single photon detection. These short gates largely reduce the number of charge carriers generated during an avalanche and, hence, the afterpulse probability, while at the same time allowing for high gate frequencies and detection rates. As the avalanche signals are very weak, standard threshold discrimination techniques are not applicable and instead, methods based on sine wave gating and filtering \cite{Namekata:06,Namekata:09,Zhang2009,zhang:76810Z,Nambu:11} or self-differencing techniques are applied \cite{yuan:041114,yuan:071101,Restelli2012}.

Here, we report on the characterization and implementation of an compact rapid gated InGaAs/InP single photon detector developed in scope of the Nano-Tera QCrypt project \cite{nanoTera2009} which aims to integrate a fast and continuous QKD system based on the coherent one-way protocol \cite{stucki-2005-87, Stucki2009, Branciard2008}. To support the high key generation rates aimed in this project, the detector is gated at 1.25~GHz and a simple but robust low-pass filtering is used to extract the weak avalanche signals. The low-pass filtering potentially bears the advantages that not only the gate frequency is efficiently filtered, but also all higher harmonics generated by non-linear gain of the amplifier or in the photo-diode. Moreover, the filtering is not restricted to a single gate frequency but works also over a wider frequency range as opposed for band stop filtering or self-differencing techniques.

This report is structured as follows: In the next section we briefly motivate and introduce the implemented gating scheme. In section~\ref{sec:Characterisation} we characterize the detector's performance in terms of detection efficiency, and noise contributions and timing jitter, and finally we provide an estimate on the performance in potential QKD scenarios.
\section{Setup}
\begin{figure*}[tbp]
\begin{minipage}[t]{\columnwidth}
\centering
\includegraphics[width=0.98\columnwidth]{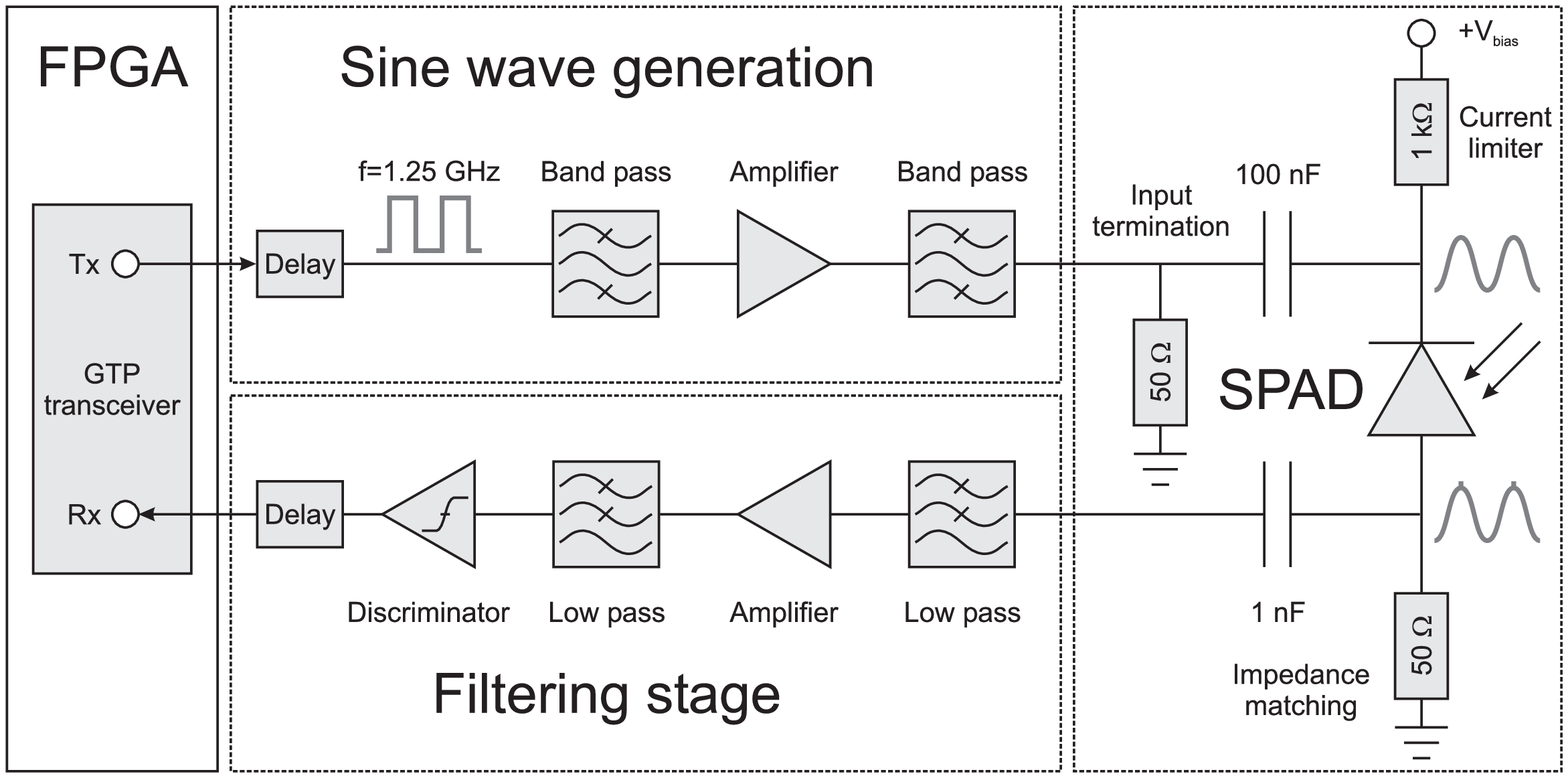}
\caption{Sketch of the implemented sine gating and low-pass filtering technique for single photon avalanche diodes. The upper part represents the sine gate generation stage, the lower part the low-pass filtering stage. The GTP transceiver interconnects the detector with an FPGA for supervision and acquisition.}
\label{fig:SineGating}
\end{minipage}
\hspace{0.1cm}
\begin{minipage}[t]{\columnwidth}
\centering
\includegraphics[width=0.81\columnwidth]{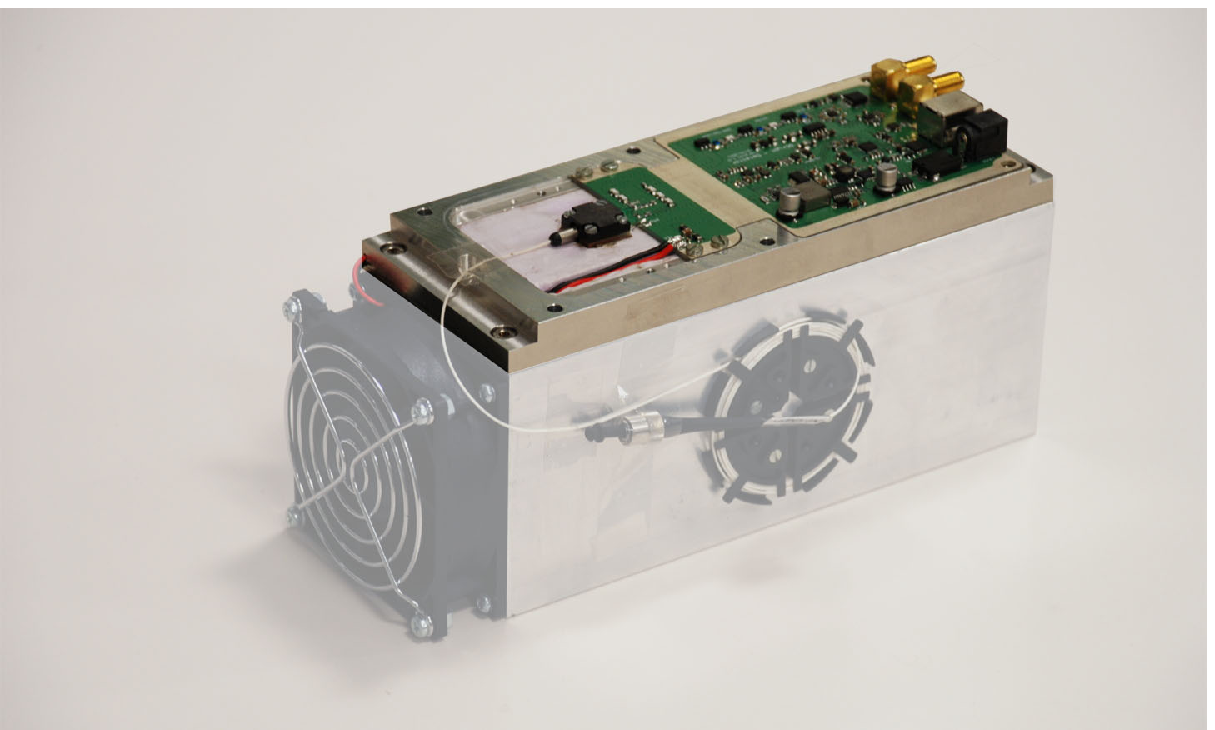}
\caption{The detector box with the fiber pigtailed avalanche diode in front and the bias and filtering electronics behind. One SMA connector is the input for the gate signals, the second the detection output. The shaded area is the cooling body.}
\label{fig:Detector}
\end{minipage}
\end{figure*}

The setup of our detection scheme based on the sine-wave gating and low-pass filtering technique is depicted in Fig.~\ref{fig:SineGating}. It comprises essentially the sine wave generation stage, the single photon avalanche diode (SPAD, JDS Uniphase) in a capacitively coupled bias-T circuit, the filtering stage and a supervising Xilinx Virtex-5 FPGA (field-programmable gate array) for generating the raw gate signals and for post-processing the detection signals. We use a fast Gigabit transceiver (Xilinx GTP) to interconnect the FPGA with the detector hardware.

The sine gates are generated by the GTP transmitter periodically transmitting square pulses with the gate frequency, which subsequently can be digitally delayed in 10~ps intervals in order to allow a temporal tuning between the gates and the incoming signals. This electrical pulse train is filtered using a sequence of band-pass filters with spectra centered around the gate frequency and a width sufficiently small to remove higher harmonic frequency components. The obtained sine signal is amplified to the required amplitude determined by the desired gate width. After a further filtering step to remove frequency sidebands potentially generated by non-linear gain in the power amplifier, the produced gate signals typically have a peak-to-peak amplitude of around 8~V. 

\begin{figure}[btp]
	\centering
	\includegraphics[width=0.95\columnwidth]{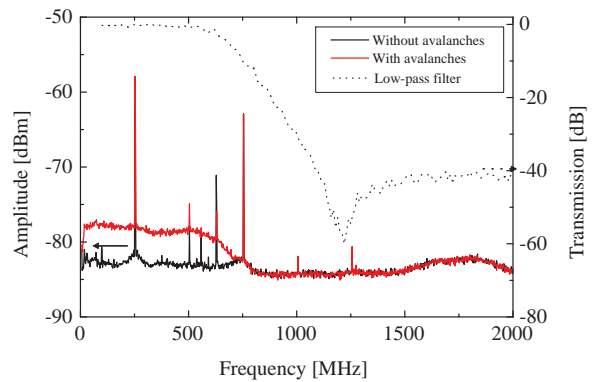}
	\caption{Dotted curve: Transmission spectrum of one of the low-pass filters. Solid curves: Spectra of the detector output after the low-pass filter stage without (black) and with (red) weak illumination. The gate  frequency at 1.25~GHz is rendered negligible by the filter stage while the avalanche signals can still be extracted from their frequency components below 1~GHz.}
 \label{fig:AvalancheSpectra}
\end{figure}

Under sine gating, the unfiltered output signal of the SPAD is largely dominated by the capacitive response of the diode to the gate frequency. To discriminate the weak avalanche signals which are orders of magnitude smaller, we process the output of the SPAD by a low noise power amplifier (Wenteq 20-3000~MHz) and a sequence of low pass filters (Crystek CLPFL-0600) which suppress the capacitive response of the diode. Each filter has a transmission spectrum as shown in Fig.~\ref{fig:AvalancheSpectra} (dotted curve). It yields a 1.25~GHz band rejection of -54~dB, and we measured more than -40~dB attenuation for higher frequencies up to 4~GHz, sufficient to suppresses the most dominant harmonic frequencies. Moreover, as the band rejection remains below -50~dB over a range of more than 100~MHz, this filtering technique does not require a precise frequency matching between the gate frequency and the filter as opposed to band stop filtering or self-differencing techniques. However, throughout the report we restrict our analysis to a gate frequency of 1.25~GHz.

After the filtering stage the avalanche signal is discriminated by a standard threshold discriminator. A final delay allows to temporally adjust the discriminator pulses with the 400~ps sampling cycles of the FPGA. We note, that in general sine gating prevents the usual approach of applying a hold-off time after each detection during which the detector is biased below breakdown and no avalanche, including due to afterpulsing, can be triggered. Instead, we post-process detections depending on their separation from the previous detection and discard all events which occur during a certain hold-off period. If not stated otherwise, we have programmed the FPGA for all the results presented here with a logic hold-off period of 10~gates corresponding to 8~ns during which we discard all detections.
\section{Characterization}\label{sec:Characterisation}
\begin{figure}[btp]
\centering
\includegraphics[width=0.9\columnwidth]{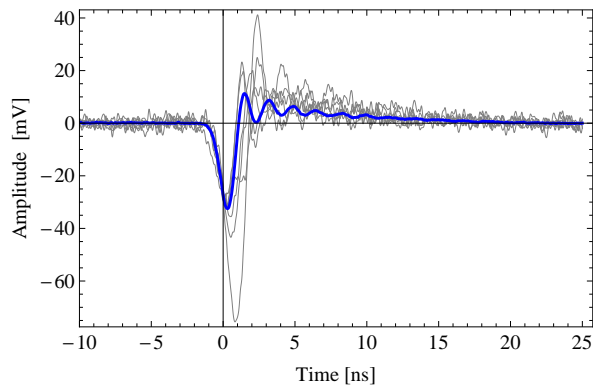}
\caption{Average avalanche signal (blue) and examples of individually captured avalanche signals (gray) after the second low-pass filter. The average amplitude of the negative peak is -32~mV, the fall time 1.8~ns.}
\label{fig:OutPulse}
\end{figure}

To characterize the performance of the detector, we use a synchronized output of the FPGA at 31.25~MHz to trigger a pulsed laser diode (PicoQuant PDL 800-B) producing short 30~ps optical pulses which are attenuated to 0.1 photons per pulse and temporally well aligned with the detector gates. 
The spectra of the detector output are shown in Fig.~\ref{fig:AvalancheSpectra} without and with weak illumination and indicate that sufficient contributions remain even by considering only frequency components below 1~GHz. In Fig.~\ref{fig:OutPulse} the average avalanche signal after the second low-pass filter captured with a 6~GHz digitizing oscilloscope (LeCroy Wavemaster 8600) is shown. It yields a clear peak of 32~mV amplitude and a falling edge of 1.8~ns length.

\begin{figure}[btp]
\centering
\includegraphics[width=0.9\columnwidth]{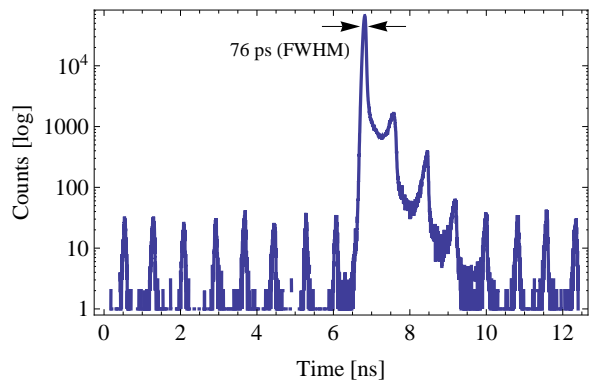}
\caption{Histogram of detection times yielding a peak width (FWHM) of 76~ps corresponding to a detector timing jitter as low as 70~ps. Around 2.4~\% of all detections appear in one of the three subsequent gates after the main peak.}
\label{fig:TCSPCHist}
\end{figure}

The gray curves in Fig.~\ref{fig:OutPulse} show examples of individual avalanche signals which exhibit fluctuations in amplitude and width. In order to characterize the impact of these fluctuating pulse shapes on the detection jitter we measured the detection time histogram using a time-correlated single photon counting module (b\&h SPC-130). In Fig.~\ref{fig:TCSPCHist} the histogram yields a major peak of width 76~ps (FWHM) from laser pulse detections. Taking the laser pulse width and jitter into account this corresponds to a very low detector timing jitter of only 70~ps.

However, after the main histogram peak, detections also occur between gates, and around 2.4~\% of all detections appear in one of the three subsequent gates after the main peak where they aren't expected. As the TCSPC module only registers the first detection per cycle, the detection in subsequent gates could not be due to afterpulsing. Additionally, we used the FPGA acquisition to analyze the correlations between subsequent detections which confirmed that these erroneous detections do not yield the correlations typically observed from afterpulsing. From this we rather identify the timing jitter induced by the low pass filters and the fluctuating amplitudes of the avalanche signals as sources of these errors. Replacing the standard discriminator by a constant-fraction discriminator (CFD) might help to reduce this effect in future.

Nevertheless, as a second possibility these errors could still be caused by afterpulsing, if the previous avalanche triggering an afterpulse is too small to be discriminated. In this case, no correlations are observed. This could happen for detections occurring at the end of a gate where the number of generated avalanche electrons is much smaller than for detections at the beginning of a gate.

\begin{figure}[btp]
\centering
\includegraphics[width=0.95\columnwidth]{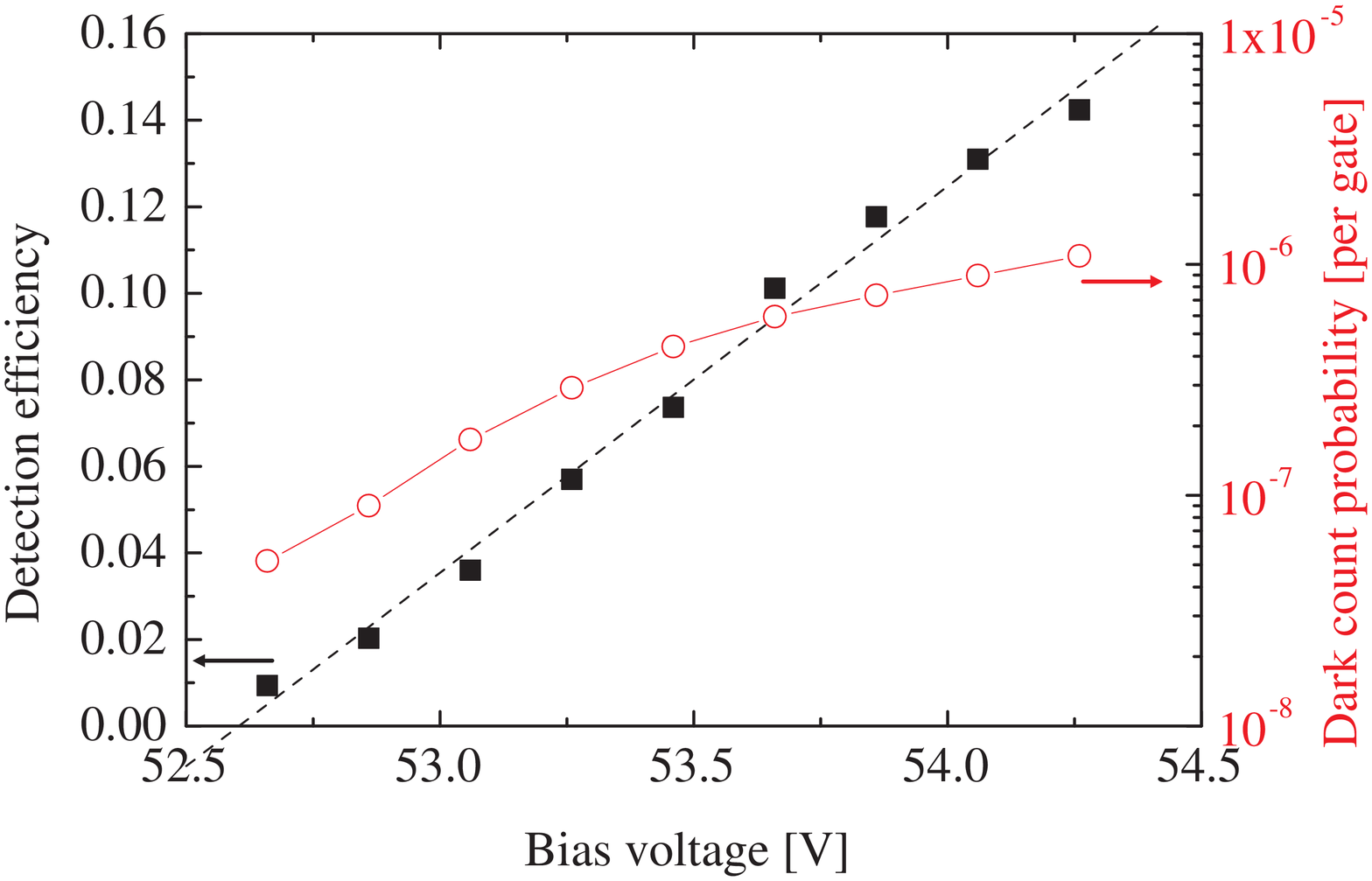}
\caption{Detection efficiency (squares) and dark count probability per gate (circles) when the bias voltage is scanned. The dashed line shows a linear fit of the detection efficiency dependence.}
\label{fig:DetEffDCBias}
\end{figure}

\begin{figure}[btp]
	\centering
	\includegraphics[width=0.95\columnwidth]{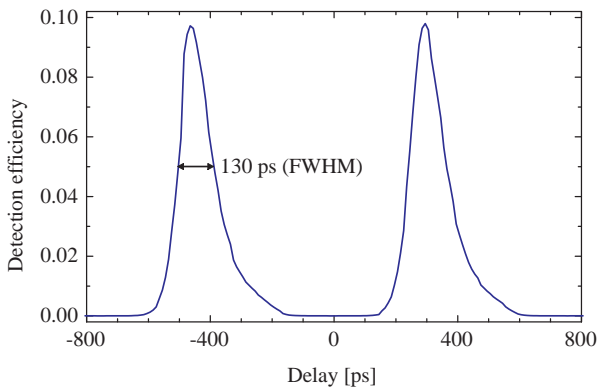}
	\caption{Detection efficiency obtained when the delay between optical pulses and the sine gate is scanned. The effective width of the gates depends on the sine amplitude and is 130~ps in the configuration shown.}
 \label{fig:GateShape}
\end{figure}

Fig.~\ref{fig:DetEffDCBias} shows the performance of the detector in terms of dark count probability per gate and detection efficiency when the temperature of the SPAD is reduced to $-43^{\circ}$C and the bias voltage is scanned. With increasing bias voltage the detection efficiency increases linearly and corresponds to 0.1 when the dark count probability per gate is $6\cdot 10^{-7}$. To measure the effective width of the gate, we change the temporal alignment between the incoming optical pulses and the detector gate by scanning the upper delay in Fig.~\ref{fig:SineGating}. From the resulting detection statistics in Fig.~\ref{fig:GateShape} we obtain the maximal detection efficiency of 0.1 in the center while the full width at half maximum gate width is as low as 130~ps. The dark count probability normalized per nanosecond hence corresponds to $5\cdot 10^{-6}$ comparable to the results obtained using the diode with a conventional low frequency gating technique.

\begin{figure}[btp]
\centering
\includegraphics[width=0.95\columnwidth]{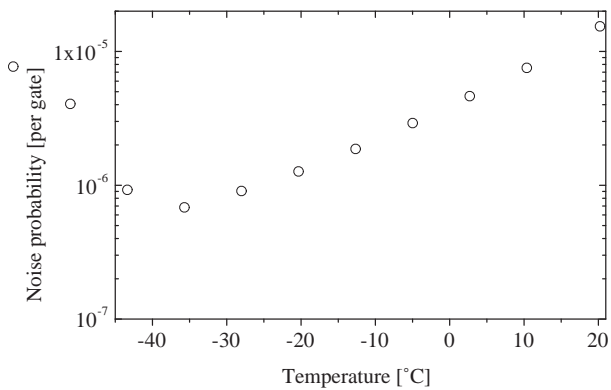}
\caption{Measured noise probability per gate as a function of detector temperature with a minimum value of only $7\cdot10^{-7}$ per gate, and a probability of $1.5\cdot10^{-5}$ per gate at $20^{\circ}$C.}
\label{fig:pDCTempDep}
\end{figure}

As mentioned earlier, the impairment due to different noise contributions depends largely on the detector temperature. In Fig.~\ref{fig:pDCTempDep} we show the dark count probability per gate, i.e. the probability of a detection when no light is impinging the detector and the temperature is varied over a range between $-45^{\circ}$C and $+20^{\circ}$C. The bias voltage is adjusted respectively to keep the excess bias constant such that the detection efficiency would constantly amount to 0.1. The minimum amount of noise corresponding to a probability of $6.8\cdot10^{-7}$ per gate is detected at a temperature of $-35^{\circ}$C while for higher temperatures the amount of thermally activated dark counts is increased.
\section{Detector performance in a QKD scenario}
Due to its very low dark count probability and timing jitter, the detection scheme presented here promises to be well suited for high rate, long distance quantum key distribution. In the following we present the results which characterize the performance of the detector in a QKD setup based on the coherent one-way protocol \cite{stucki-2005-87}, but equivalent results in terms of detection rate and quantum bit error rate (QBER) should be achievable with any other QKD setup.

In the COW implementation, Alice tailors the output of a continuous coherent laser to produce pulse-position modulated (PPM \cite{Hamkins2007}) quantum states, where one bit of information is encoded in the position of an optical pulse out of two possible time-bins per bit. For pulse shaping we use an intensity modulator (Photline MXER-LN) with a measured extinction ratio of 25~dB, the bit frequency is 625~Mbps (mega bits per second). At the receiver side, Bob decodes the information by analyzing the detection times with his FPGA. Errors in this detection schemes correspond to events registered in the wrong of two possible time-bins of 400~ps width. All detections outside the 400~ps wide time-bins are discarded and do not contribute to the detection or error rates.

\begin{figure}[btp]
	\centering
	\includegraphics[width=0.95\columnwidth]{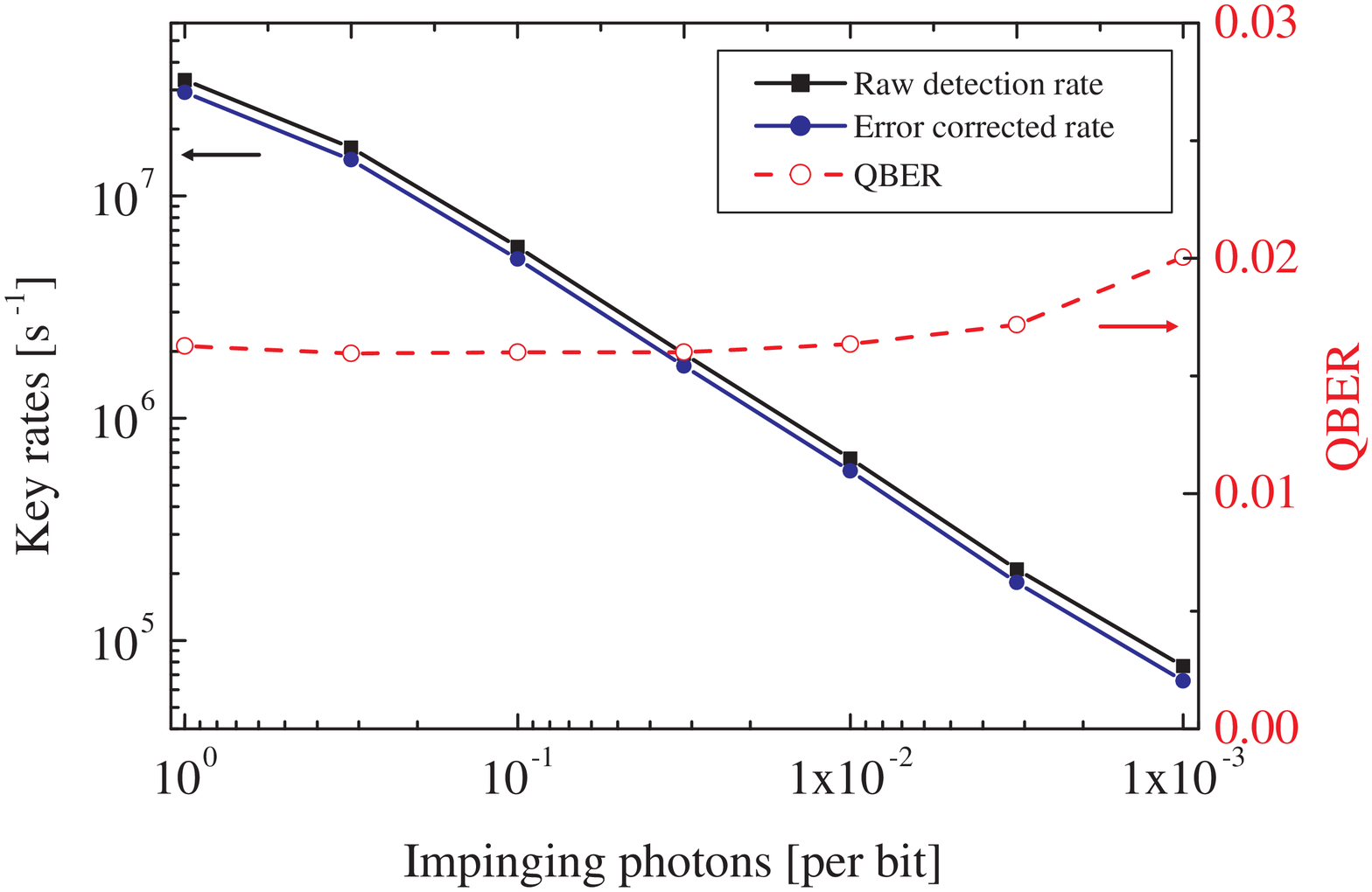}
	\caption{Detection rates for the sine gated APD in a QKD scenario. From the measured raw detection rates and QBER we estimate a COW secret key rate \cite{Branciard2008} larger 1~Mbps up to 4~dB fiber losses (corresponding to 20~km standard fiber length).}
 \label{fig:SKRs}
\end{figure}

First, we estimate the performance of the detector for QKD over different fiber lengths. As the average number of photons per bit reaching the detector does not only depend on the fiber transmission, but also on the number of photons sent by Alice, and, hence, on the implemented QKD protocol, the results presented in Fig.~\ref{fig:SKRs} are given as a function of the number of impinging photons per bit. Here, the detector temperature is $-43^{\circ}$C and the dc-bias voltage 53.5~V. For an average of 1~photon per bit the detection rate is 33~Mbps, corresponding well to the expected rate of 31~Mbps when a Poissonian photon number distribution and reduction due to the 8~ns hold-off time is taken into account. With lower photon numbers, the raw detection rate decreases linearly as expected.

The QBER in Fig.~\ref{fig:SKRs} yields a value as low as 2.0~\% for 0.001 impinging photons per bit, matching the expected value given by the detector noise. In a standard BB84 setup without decoy states this impinging photon number would correspond to a transmission over 75~km standard fiber, for a COW implementation it would correspond to 115~km. However, at higher photon numbers per bit (or shorter fiber lengths), the QBER yields a minimum value of 1.6~\% which is above the expected lower limit of 0.2~\% given by the extinction ratio (25~dB) of Alice' intensity modulator preparing the states. As discussed in the previous section, the erroneous detections were not correlated with previous detections and for the moment we can't distinguish with certainty whether they stem from timing jitter or afterpulsing. However, in Fig.~\ref{fig:TCSPCHist} around 2.4~\% of all detections appeared in one of the three gates after the main detection peak, which together with errors due to the extinction ratio of Alice' intensity modulator explain an error rate of at least 1.4~\%. Taking the measured QBER into account, we also show the expected rates corresponding to the situation after error correction. For the COW protocol (assuming individual attacks), the performance of our detector would allow to generate more than 1~Mbps (mega bit per second) of secret keys up to a fiber attenuation of 4~dB, corresponding to a standard fiber with attenuation coefficient 0.2~dB/km of 20~km length.

\begin{figure}[btp]
	\centering
	\includegraphics[width=0.95\columnwidth]{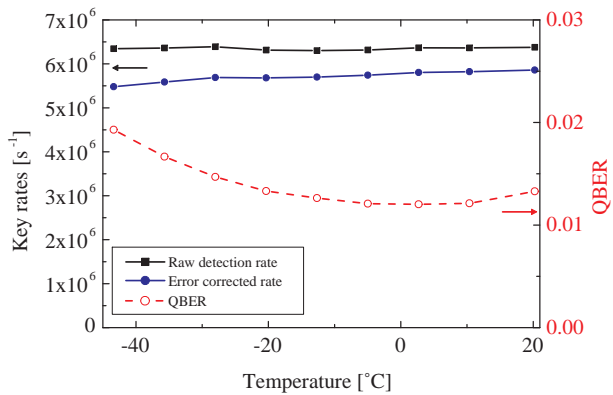}
	\caption{Measured detection rates and quantum bit error rates for different detector temperatures and the corresponding key rate after error correction. The incoming photon intensity of 0.1 photons per pulse corresponds to the optimal QKD configuration over 25~km fiber length (using standard BB84 or COW). Even at room temperature the detector allows for high rate QKD.}
 \label{fig:rECTemp}
\end{figure}

We have also tested the detector performance for QKD at different temperatures. At a bias voltage of 53.5~V the laser pulses were attenuated down to 0.1~photons per pulse at the detector input, corresponding to the optimum configuration over 25 km fiber length with 0.3 photons per pulse. The results in Fig.~\ref{fig:rECTemp} show that for this fiber lengths the detector allows for high rate QKD even at room temperature of $20~^{\circ}$C. Although this result stems merely from the very short gate durations at high gate frequency which reduces the amount of detected noise per gate, it suggests that the size and complexity of current single photon detectors could largely be reduced as no bulky cooling mechanics is required (see Fig.~\ref{fig:Detector}). This might allow for reducing the mechanical dimensions of future QKD platforms as developed in the QCrypt project \cite{nanoTera2009} and might even pave the way for the development of small handheld size QKD devices.

\begin{figure}[btp]
	\centering
	\includegraphics[width=0.95\columnwidth]{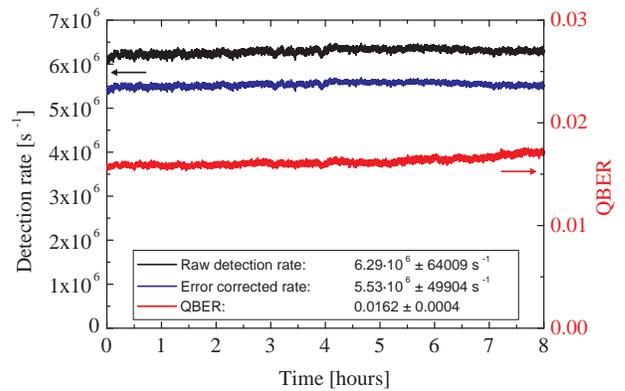}
	\caption{Measurement results showing the stability of the detector over a period of 8~h operation.}
 \label{fig:SKRsTime}
\end{figure}

As stability and reliability of the detection hardware is a major requirement for robust long-term quantum key distribution, we have run the detector over several hours and monitored the detection statistics. As the curves in Fig.~\ref{fig:SKRsTime} show, the detection and error rates remain constant over a measurement period of 8 hours without any further stabilization. However, as we expect temporal alignment drifts between the optical pulses and the detector gates for long fiber lengths, the final QKD setup will comprise a continuous tracking system to actively compensate for temperature induced fiber length variations.
\section{Conclusions}
We have presented the results of an implementation and characterization of a 1.25~GHz gated InGaAs/InP single photon avalanche diode with simple low-pass filtering technique suitable for fast QKD operation. We have demonstrated that our sine gating and filtering technique largely limits the amount of detected noise as the gate width is as short as 130~ps. At a detection efficiency of 0.1 we obtain a dark count probability as low as $7\cdot10^{-7}$ per gate and a timing jitter of 70~ps. The detector works remarkably well over a wide temperature range and allows for QKD over more than 25~km even at room temperature. As the low-pass filtering technique for sine gating does not require a precise adaption to the gate frequency and is very robust against environmental temperature fluctuations or mechanical stresses, the results presented here might be used to reduce the size and complexity of fast single photon detectors at telecom wavelength and of quantum communication systems employing them.
\section{Acknowledgements}
We gratefully acknowledge the fruitful input and discussions with Herv\'e Eus\`ebe and Gabriel Bernasconi from Hepia Geneva, as well as Matthieu Legr\'e from ID Quantique SA. This research was financially supported by the Swiss Nano-Tera project QCRYPT and the National Center of Competence in Research QSIT.
%

\begin{thebibliography}{28}%
\makeatletter
\providecommand \@ifxundefined [1]{%
 \@ifx{#1\undefined}
}%
\providecommand \@ifnum [1]{%
 \ifnum #1\expandafter \@firstoftwo
 \else \expandafter \@secondoftwo
 \fi
}%
\providecommand \@ifx [1]{%
 \ifx #1\expandafter \@firstoftwo
 \else \expandafter \@secondoftwo
 \fi
}%
\providecommand \natexlab [1]{#1}%
\providecommand \enquote  [1]{``#1''}%
\providecommand \bibnamefont  [1]{#1}%
\providecommand \bibfnamefont [1]{#1}%
\providecommand \citenamefont [1]{#1}%
\providecommand \href@noop [0]{\@secondoftwo}%
\providecommand \href [0]{\begingroup \@sanitize@url \@href}%
\providecommand \@href[1]{\@@startlink{#1}\@@href}%
\providecommand \@@href[1]{\endgroup#1\@@endlink}%
\providecommand \@sanitize@url [0]{\catcode `\\12\catcode `\$12\catcode
  `\&12\catcode `\#12\catcode `\^12\catcode `\_12\catcode `\%12\relax}%
\providecommand \@@startlink[1]{}%
\providecommand \@@endlink[0]{}%
\providecommand \url  [0]{\begingroup\@sanitize@url \@url }%
\providecommand \@url [1]{\endgroup\@href {#1}{\urlprefix }}%
\providecommand \urlprefix  [0]{URL }%
\providecommand \Eprint [0]{\href }%
\providecommand \doibase [0]{http://dx.doi.org/}%
\providecommand \selectlanguage [0]{\@gobble}%
\providecommand \bibinfo  [0]{\@secondoftwo}%
\providecommand \bibfield  [0]{\@secondoftwo}%
\providecommand \translation [1]{[#1]}%
\providecommand \BibitemOpen [0]{}%
\providecommand \bibitemStop [0]{}%
\providecommand \bibitemNoStop [0]{.\EOS\space}%
\providecommand \EOS [0]{\spacefactor3000\relax}%
\providecommand \BibitemShut  [1]{\csname bibitem#1\endcsname}%
\let\auto@bib@innerbib\@empty
\bibitem [{\citenamefont {Bennett}\ and\ \citenamefont
  {Brassard}(1984)}]{Bennett84}%
  \BibitemOpen
  \bibfield  {author} {\bibinfo {author} {\bibfnamefont {C.~H.}\ \bibnamefont
  {Bennett}}\ and\ \bibinfo {author} {\bibfnamefont {G.}~\bibnamefont
  {Brassard}},\ }in\ \href@noop {} {\emph {\bibinfo {booktitle} {Int.
  Conference on Computers, Systems and Signal Processing}}}\ (\bibinfo {year}
  {1984})\ pp.\ \bibinfo {pages} {175--179}\BibitemShut {NoStop}%
\bibitem [{\citenamefont {Gisin}\ \emph {et~al.}(2002)\citenamefont {Gisin},
  \citenamefont {Ribordy}, \citenamefont {Tittel},\ and\ \citenamefont
  {Zbinden}}]{Gisin2002b}%
  \BibitemOpen
  \bibfield  {author} {\bibinfo {author} {\bibfnamefont {N.}~\bibnamefont
  {Gisin}}, \bibinfo {author} {\bibfnamefont {G.}~\bibnamefont {Ribordy}},
  \bibinfo {author} {\bibfnamefont {W.}~\bibnamefont {Tittel}}, \ and\ \bibinfo
  {author} {\bibfnamefont {H.}~\bibnamefont {Zbinden}},\ }\href@noop {}
  {\bibfield  {journal} {\bibinfo  {journal} {Reviews of Modern Physics}\
  }\textbf {\bibinfo {volume} {74}},\ \bibinfo {pages} {145} (\bibinfo {year}
  {2002})}\BibitemShut {NoStop}%
\bibitem [{\citenamefont {Eraerds}\ \emph {et~al.}(2010)\citenamefont
  {Eraerds}, \citenamefont {Legre}, \citenamefont {Zhang}, \citenamefont
  {Zbinden},\ and\ \citenamefont {Gisin}}]{Eraerds2010a}%
  \BibitemOpen
  \bibfield  {author} {\bibinfo {author} {\bibfnamefont {P.}~\bibnamefont
  {Eraerds}}, \bibinfo {author} {\bibfnamefont {M.}~\bibnamefont {Legre}},
  \bibinfo {author} {\bibfnamefont {J.}~\bibnamefont {Zhang}}, \bibinfo
  {author} {\bibfnamefont {H.}~\bibnamefont {Zbinden}}, \ and\ \bibinfo
  {author} {\bibfnamefont {N.}~\bibnamefont {Gisin}},\ }\href {\doibase
  10.1109/JLT.2009.2039635} {\bibfield  {journal} {\bibinfo  {journal}
  {IEEE/OSA Journal of Lightwave Technology}\ }\textbf {\bibinfo {volume}
  {28}},\ \bibinfo {pages} {952} (\bibinfo {year} {2010})}\BibitemShut
  {NoStop}%
\bibitem [{\citenamefont {Healey}\ and\ \citenamefont
  {Hensel}(1980)}]{Healey1980}%
  \BibitemOpen
  \bibfield  {author} {\bibinfo {author} {\bibfnamefont {P.}~\bibnamefont
  {Healey}}\ and\ \bibinfo {author} {\bibfnamefont {P.}~\bibnamefont
  {Hensel}},\ }\href {\doibase 10.1049/el:19800438} {\bibfield  {journal}
  {\bibinfo  {journal} {Electron. Lett}\ }\textbf {\bibinfo {volume} {16}},\
  \bibinfo {pages} {631} (\bibinfo {year} {1980})}\BibitemShut {NoStop}%
\bibitem [{\citenamefont {Priedhorsky}, \citenamefont {Smith},\ and\
  \citenamefont {Ho}(1996)}]{Priedhorsky:96}%
  \BibitemOpen
  \bibfield  {author} {\bibinfo {author} {\bibfnamefont {W.~C.}\ \bibnamefont
  {Priedhorsky}}, \bibinfo {author} {\bibfnamefont {R.~C.}\ \bibnamefont
  {Smith}}, \ and\ \bibinfo {author} {\bibfnamefont {C.}~\bibnamefont {Ho}},\
  }\href {\doibase 10.1364/AO.35.000441} {\bibfield  {journal} {\bibinfo
  {journal} {Appl. Opt.}\ }\textbf {\bibinfo {volume} {35}},\ \bibinfo {pages}
  {441} (\bibinfo {year} {1996})}\BibitemShut {NoStop}%
\bibitem [{\citenamefont {Warburton}\ \emph {et~al.}(2007)\citenamefont
  {Warburton}, \citenamefont {McCarthy}, \citenamefont {Wallace}, \citenamefont
  {Hernandez-Marin}, \citenamefont {Hadfield}, \citenamefont {Nam},\ and\
  \citenamefont {Buller}}]{Warburton:07}%
  \BibitemOpen
  \bibfield  {author} {\bibinfo {author} {\bibfnamefont {R.~E.}\ \bibnamefont
  {Warburton}}, \bibinfo {author} {\bibfnamefont {A.}~\bibnamefont {McCarthy}},
  \bibinfo {author} {\bibfnamefont {A.~M.}\ \bibnamefont {Wallace}}, \bibinfo
  {author} {\bibfnamefont {S.}~\bibnamefont {Hernandez-Marin}}, \bibinfo
  {author} {\bibfnamefont {R.~H.}\ \bibnamefont {Hadfield}}, \bibinfo {author}
  {\bibfnamefont {S.~W.}\ \bibnamefont {Nam}}, \ and\ \bibinfo {author}
  {\bibfnamefont {G.~S.}\ \bibnamefont {Buller}},\ }\href {\doibase
  10.1364/OL.32.002266} {\bibfield  {journal} {\bibinfo  {journal} {Opt.
  Lett.}\ }\textbf {\bibinfo {volume} {32}},\ \bibinfo {pages} {2266} (\bibinfo
  {year} {2007})}\BibitemShut {NoStop}%
\bibitem [{\citenamefont {Ren}\ \emph {et~al.}(2011)\citenamefont {Ren},
  \citenamefont {Gu}, \citenamefont {Liang}, \citenamefont {Kong},
  \citenamefont {Wu}, \citenamefont {Wu},\ and\ \citenamefont {Zeng}}]{Ren:11}%
  \BibitemOpen
  \bibfield  {author} {\bibinfo {author} {\bibfnamefont {M.}~\bibnamefont
  {Ren}}, \bibinfo {author} {\bibfnamefont {X.}~\bibnamefont {Gu}}, \bibinfo
  {author} {\bibfnamefont {Y.}~\bibnamefont {Liang}}, \bibinfo {author}
  {\bibfnamefont {W.}~\bibnamefont {Kong}}, \bibinfo {author} {\bibfnamefont
  {E.}~\bibnamefont {Wu}}, \bibinfo {author} {\bibfnamefont {G.}~\bibnamefont
  {Wu}}, \ and\ \bibinfo {author} {\bibfnamefont {H.}~\bibnamefont {Zeng}},\
  }\href {\doibase 10.1364/OE.19.013497} {\bibfield  {journal} {\bibinfo
  {journal} {Opt. Express}\ }\textbf {\bibinfo {volume} {19}},\ \bibinfo
  {pages} {13497} (\bibinfo {year} {2011})}\BibitemShut {NoStop}%
\bibitem [{\citenamefont {Gol'tsman}\ \emph {et~al.}(2001)\citenamefont
  {Gol'tsman}, \citenamefont {Okunev}, \citenamefont {Chulkova}, \citenamefont
  {Lipatov}, \citenamefont {Semenov}, \citenamefont {Smirnov}, \citenamefont
  {Voronov}, \citenamefont {Dzardanov}, \citenamefont {Williams},\ and\
  \citenamefont {Sobolewski}}]{goltsman:705}%
  \BibitemOpen
  \bibfield  {author} {\bibinfo {author} {\bibfnamefont {G.~N.}\ \bibnamefont
  {Gol'tsman}}, \bibinfo {author} {\bibfnamefont {O.}~\bibnamefont {Okunev}},
  \bibinfo {author} {\bibfnamefont {G.}~\bibnamefont {Chulkova}}, \bibinfo
  {author} {\bibfnamefont {A.}~\bibnamefont {Lipatov}}, \bibinfo {author}
  {\bibfnamefont {A.}~\bibnamefont {Semenov}}, \bibinfo {author} {\bibfnamefont
  {K.}~\bibnamefont {Smirnov}}, \bibinfo {author} {\bibfnamefont
  {B.}~\bibnamefont {Voronov}}, \bibinfo {author} {\bibfnamefont
  {A.}~\bibnamefont {Dzardanov}}, \bibinfo {author} {\bibfnamefont
  {C.}~\bibnamefont {Williams}}, \ and\ \bibinfo {author} {\bibfnamefont
  {R.}~\bibnamefont {Sobolewski}},\ }\href {\doibase 10.1063/1.1388868}
  {\bibfield  {journal} {\bibinfo  {journal} {Applied Physics Letters}\
  }\textbf {\bibinfo {volume} {79}},\ \bibinfo {pages} {705} (\bibinfo {year}
  {2001})}\BibitemShut {NoStop}%
\bibitem [{\citenamefont {Korneev}\ \emph {et~al.}(2007)\citenamefont
  {Korneev}, \citenamefont {Vachtomin}, \citenamefont {Minaeva}, \citenamefont
  {Divochiy}, \citenamefont {Smirnov}, \citenamefont {Okunev}, \citenamefont
  {Gol'tsman}, \citenamefont {Zinoni}, \citenamefont {Chauvin}, \citenamefont
  {Balet}, \citenamefont {Marsili}, \citenamefont {Bitauld}, \citenamefont
  {Alloing}, \citenamefont {Li}, \citenamefont {Fiore}, \citenamefont {Lunghi},
  \citenamefont {Gerardino}, \citenamefont {Halder}, \citenamefont {Jorel},\
  and\ \citenamefont {Zbinden}}]{SSPD2007}%
  \BibitemOpen
  \bibfield  {author} {\bibinfo {author} {\bibfnamefont {A.}~\bibnamefont
  {Korneev}}, \bibinfo {author} {\bibfnamefont {Y.}~\bibnamefont {Vachtomin}},
  \bibinfo {author} {\bibfnamefont {O.}~\bibnamefont {Minaeva}}, \bibinfo
  {author} {\bibfnamefont {A.}~\bibnamefont {Divochiy}}, \bibinfo {author}
  {\bibfnamefont {K.}~\bibnamefont {Smirnov}}, \bibinfo {author} {\bibfnamefont
  {O.}~\bibnamefont {Okunev}}, \bibinfo {author} {\bibfnamefont
  {G.}~\bibnamefont {Gol'tsman}}, \bibinfo {author} {\bibfnamefont
  {C.}~\bibnamefont {Zinoni}}, \bibinfo {author} {\bibfnamefont
  {N.}~\bibnamefont {Chauvin}}, \bibinfo {author} {\bibfnamefont
  {L.}~\bibnamefont {Balet}}, \bibinfo {author} {\bibfnamefont
  {F.}~\bibnamefont {Marsili}}, \bibinfo {author} {\bibfnamefont
  {D.}~\bibnamefont {Bitauld}}, \bibinfo {author} {\bibfnamefont
  {B.}~\bibnamefont {Alloing}}, \bibinfo {author} {\bibfnamefont
  {L.}~\bibnamefont {Li}}, \bibinfo {author} {\bibfnamefont {A.}~\bibnamefont
  {Fiore}}, \bibinfo {author} {\bibfnamefont {L.}~\bibnamefont {Lunghi}},
  \bibinfo {author} {\bibfnamefont {A.}~\bibnamefont {Gerardino}}, \bibinfo
  {author} {\bibfnamefont {M.}~\bibnamefont {Halder}}, \bibinfo {author}
  {\bibfnamefont {C.}~\bibnamefont {Jorel}}, \ and\ \bibinfo {author}
  {\bibfnamefont {H.}~\bibnamefont {Zbinden}},\ }\href {\doibase
  10.1109/JSTQE.2007.903856} {\bibfield  {journal} {\bibinfo  {journal} {IEEE
  Selected Topics in Quantum Electronics}\ }\textbf {\bibinfo {volume} {13}},\
  \bibinfo {pages} {944} (\bibinfo {year} {2007})}\BibitemShut {NoStop}%
\bibitem [{\citenamefont {Stucki}\ \emph
  {et~al.}(2009{\natexlab{a}})\citenamefont {Stucki}, \citenamefont {Walenta},
  \citenamefont {Vannel}, \citenamefont {Thew}, \citenamefont {Gisin},
  \citenamefont {Zbinden}, \citenamefont {Gray}, \citenamefont {Towery},\ and\
  \citenamefont {Ten}}]{Stucki2009a}%
  \BibitemOpen
  \bibfield  {author} {\bibinfo {author} {\bibfnamefont {D.}~\bibnamefont
  {Stucki}}, \bibinfo {author} {\bibfnamefont {N.}~\bibnamefont {Walenta}},
  \bibinfo {author} {\bibfnamefont {F.}~\bibnamefont {Vannel}}, \bibinfo
  {author} {\bibfnamefont {R.~T.}\ \bibnamefont {Thew}}, \bibinfo {author}
  {\bibfnamefont {N.}~\bibnamefont {Gisin}}, \bibinfo {author} {\bibfnamefont
  {H.}~\bibnamefont {Zbinden}}, \bibinfo {author} {\bibfnamefont
  {S.}~\bibnamefont {Gray}}, \bibinfo {author} {\bibfnamefont {C.~R.}\
  \bibnamefont {Towery}}, \ and\ \bibinfo {author} {\bibfnamefont
  {S.}~\bibnamefont {Ten}},\ }\href {\doibase 10.1088/1367-2630/11/7/075003}
  {\bibfield  {journal} {\bibinfo  {journal} {New Journal of Physics}\ }\textbf
  {\bibinfo {volume} {11}},\ \bibinfo {pages} {075003} (\bibinfo {year}
  {2009}{\natexlab{a}})}\BibitemShut {NoStop}%
\bibitem [{\citenamefont {Thew}\ \emph {et~al.}(2006)\citenamefont {Thew},
  \citenamefont {Tanzilli}, \citenamefont {Krainer}, \citenamefont {Zeller},
  \citenamefont {Rochas}, \citenamefont {Rech}, \citenamefont {Cova},
  \citenamefont {Zbinden},\ and\ \citenamefont {Gisin}}]{Thew2006}%
  \BibitemOpen
  \bibfield  {author} {\bibinfo {author} {\bibfnamefont {R.~T.}\ \bibnamefont
  {Thew}}, \bibinfo {author} {\bibfnamefont {S.}~\bibnamefont {Tanzilli}},
  \bibinfo {author} {\bibfnamefont {L.}~\bibnamefont {Krainer}}, \bibinfo
  {author} {\bibfnamefont {S.~C.}\ \bibnamefont {Zeller}}, \bibinfo {author}
  {\bibfnamefont {A.}~\bibnamefont {Rochas}}, \bibinfo {author} {\bibfnamefont
  {I.}~\bibnamefont {Rech}}, \bibinfo {author} {\bibfnamefont {S.}~\bibnamefont
  {Cova}}, \bibinfo {author} {\bibfnamefont {H.}~\bibnamefont {Zbinden}}, \
  and\ \bibinfo {author} {\bibfnamefont {N.}~\bibnamefont {Gisin}},\ }\href
  {\doibase 10.1088/1367-2630/8/3/032} {\bibfield  {journal} {\bibinfo
  {journal} {New Journal of Physics}\ }\textbf {\bibinfo {volume} {8}},\
  \bibinfo {pages} {32} (\bibinfo {year} {2006})}\BibitemShut {NoStop}%
\bibitem [{\citenamefont {Thew}\ \emph {et~al.}(2009)\citenamefont {Thew},
  \citenamefont {Curtz}, \citenamefont {Eraerds}, \citenamefont {Walenta},
  \citenamefont {Gautier}, \citenamefont {Koller}, \citenamefont {Zhang},
  \citenamefont {Gisin},\ and\ \citenamefont {Zbinden}}]{Thew2009}%
  \BibitemOpen
  \bibfield  {author} {\bibinfo {author} {\bibfnamefont {R.~T.}\ \bibnamefont
  {Thew}}, \bibinfo {author} {\bibfnamefont {N.}~\bibnamefont {Curtz}},
  \bibinfo {author} {\bibfnamefont {P.}~\bibnamefont {Eraerds}}, \bibinfo
  {author} {\bibfnamefont {N.}~\bibnamefont {Walenta}}, \bibinfo {author}
  {\bibfnamefont {J.~D.}\ \bibnamefont {Gautier}}, \bibinfo {author}
  {\bibfnamefont {E.}~\bibnamefont {Koller}}, \bibinfo {author} {\bibfnamefont
  {J.}~\bibnamefont {Zhang}}, \bibinfo {author} {\bibfnamefont
  {N.}~\bibnamefont {Gisin}}, \ and\ \bibinfo {author} {\bibfnamefont
  {H.}~\bibnamefont {Zbinden}},\ }\href {\doibase 10.1016/j.nima.2009.05.031}
  {\bibfield  {journal} {\bibinfo  {journal} {Nuclear Instruments \& Methods}\
  }\textbf {\bibinfo {volume} {610}},\ \bibinfo {pages} {16} (\bibinfo {year}
  {2009})}\BibitemShut {NoStop}%
\bibitem [{\citenamefont {Itzler}\ \emph {et~al.}(2011)\citenamefont {Itzler},
  \citenamefont {Jiang}, \citenamefont {Entwistle}, \citenamefont {Slomkowski},
  \citenamefont {Tosi}, \citenamefont {Acerbi}, \citenamefont {Zappa},\ and\
  \citenamefont {Cova}}]{Itzler2011}%
  \BibitemOpen
  \bibfield  {author} {\bibinfo {author} {\bibfnamefont {M.~A.}\ \bibnamefont
  {Itzler}}, \bibinfo {author} {\bibfnamefont {X.}~\bibnamefont {Jiang}},
  \bibinfo {author} {\bibfnamefont {M.}~\bibnamefont {Entwistle}}, \bibinfo
  {author} {\bibfnamefont {K.}~\bibnamefont {Slomkowski}}, \bibinfo {author}
  {\bibfnamefont {A.}~\bibnamefont {Tosi}}, \bibinfo {author} {\bibfnamefont
  {F.}~\bibnamefont {Acerbi}}, \bibinfo {author} {\bibfnamefont
  {F.}~\bibnamefont {Zappa}}, \ and\ \bibinfo {author} {\bibfnamefont
  {S.}~\bibnamefont {Cova}},\ }\href {\doibase 10.1080/09500340.2010.547262}
  {\bibfield  {journal} {\bibinfo  {journal} {Journal of Modern Optics}\
  }\textbf {\bibinfo {volume} {58}},\ \bibinfo {pages} {174} (\bibinfo {year}
  {2011})}\BibitemShut {NoStop}%
\bibitem [{\citenamefont {Cova}, \citenamefont {Lacaita},\ and\ \citenamefont
  {Ripamonti}(1991)}]{COVA91}%
  \BibitemOpen
  \bibfield  {author} {\bibinfo {author} {\bibfnamefont {S.}~\bibnamefont
  {Cova}}, \bibinfo {author} {\bibfnamefont {A.}~\bibnamefont {Lacaita}}, \
  and\ \bibinfo {author} {\bibfnamefont {G.}~\bibnamefont {Ripamonti}},\
  }\href@noop {} {\bibfield  {journal} {\bibinfo  {journal} {IEEE Electron
  Device Letters}\ }\textbf {\bibinfo {volume} {12}},\ \bibinfo {pages} {685}
  (\bibinfo {year} {1991})}\BibitemShut {NoStop}%
\bibitem [{\citenamefont {Kindt}()}]{Kindt99}%
  \BibitemOpen
  \bibfield  {author} {\bibinfo {author} {\bibfnamefont {W.~J.}\ \bibnamefont
  {Kindt}},\ }\emph {\bibinfo {title} {Geiger mode avalanche photodiode arrays:
  For spatially resolved single photon counting}},\ \href@noop {} {Ph.D.
  thesis},\ \bibinfo  {school} {University of Delft}\BibitemShut {NoStop}%
\bibitem [{\citenamefont {Namekata}, \citenamefont {Sasamori},\ and\
  \citenamefont {Inoue}(2006)}]{Namekata:06}%
  \BibitemOpen
  \bibfield  {author} {\bibinfo {author} {\bibfnamefont {N.}~\bibnamefont
  {Namekata}}, \bibinfo {author} {\bibfnamefont {S.}~\bibnamefont {Sasamori}},
  \ and\ \bibinfo {author} {\bibfnamefont {S.}~\bibnamefont {Inoue}},\ }\href
  {\doibase 10.1364/OE.14.010043} {\bibfield  {journal} {\bibinfo  {journal}
  {Opt. Express}\ }\textbf {\bibinfo {volume} {14}},\ \bibinfo {pages} {10043}
  (\bibinfo {year} {2006})}\BibitemShut {NoStop}%
\bibitem [{\citenamefont {Namekata}, \citenamefont {Adachi},\ and\
  \citenamefont {Inoue}(2009)}]{Namekata:09}%
  \BibitemOpen
  \bibfield  {author} {\bibinfo {author} {\bibfnamefont {N.}~\bibnamefont
  {Namekata}}, \bibinfo {author} {\bibfnamefont {S.}~\bibnamefont {Adachi}}, \
  and\ \bibinfo {author} {\bibfnamefont {S.}~\bibnamefont {Inoue}},\ }\href
  {\doibase 10.1364/OE.17.006275} {\bibfield  {journal} {\bibinfo  {journal}
  {Opt. Express}\ }\textbf {\bibinfo {volume} {17}},\ \bibinfo {pages} {6275}
  (\bibinfo {year} {2009})}\BibitemShut {NoStop}%
\bibitem [{\citenamefont {Zhang}\ \emph {et~al.}(2009)\citenamefont {Zhang},
  \citenamefont {Thew}, \citenamefont {Barreiro},\ and\ \citenamefont
  {Zbinden}}]{Zhang2009}%
  \BibitemOpen
  \bibfield  {author} {\bibinfo {author} {\bibfnamefont {J.}~\bibnamefont
  {Zhang}}, \bibinfo {author} {\bibfnamefont {R.}~\bibnamefont {Thew}},
  \bibinfo {author} {\bibfnamefont {C.}~\bibnamefont {Barreiro}}, \ and\
  \bibinfo {author} {\bibfnamefont {H.}~\bibnamefont {Zbinden}},\ }\href
  {\doibase 10.1063/1.3223576} {\bibfield  {journal} {\bibinfo  {journal}
  {Applied Physics Letters}\ }\textbf {\bibinfo {volume} {95}},\ \bibinfo
  {pages} {091103} (\bibinfo {year} {2009})}\BibitemShut {NoStop}%
\bibitem [{\citenamefont {Zhang}\ \emph {et~al.}(2010)\citenamefont {Zhang},
  \citenamefont {Eraerds}, \citenamefont {Walenta}, \citenamefont {Barreiro},
  \citenamefont {Thew},\ and\ \citenamefont {Zbinden}}]{zhang:76810Z}%
  \BibitemOpen
  \bibfield  {author} {\bibinfo {author} {\bibfnamefont {J.}~\bibnamefont
  {Zhang}}, \bibinfo {author} {\bibfnamefont {P.}~\bibnamefont {Eraerds}},
  \bibinfo {author} {\bibfnamefont {N.}~\bibnamefont {Walenta}}, \bibinfo
  {author} {\bibfnamefont {C.}~\bibnamefont {Barreiro}}, \bibinfo {author}
  {\bibfnamefont {R.}~\bibnamefont {Thew}}, \ and\ \bibinfo {author}
  {\bibfnamefont {H.}~\bibnamefont {Zbinden}}\ }(\bibinfo  {publisher} {SPIE},\
  \bibinfo {year} {2010})\ p.\ \bibinfo {pages} {76810Z}\BibitemShut {NoStop}%
\bibitem [{\citenamefont {Nambu}\ \emph {et~al.}(2011)\citenamefont {Nambu},
  \citenamefont {Takahashi}, \citenamefont {Yoshino}, \citenamefont {Tanaka},
  \citenamefont {Fujiwara}, \citenamefont {Sasaki}, \citenamefont {Tajima},
  \citenamefont {Yorozu},\ and\ \citenamefont {Tomita}}]{Nambu:11}%
  \BibitemOpen
  \bibfield  {author} {\bibinfo {author} {\bibfnamefont {Y.}~\bibnamefont
  {Nambu}}, \bibinfo {author} {\bibfnamefont {S.}~\bibnamefont {Takahashi}},
  \bibinfo {author} {\bibfnamefont {K.}~\bibnamefont {Yoshino}}, \bibinfo
  {author} {\bibfnamefont {A.}~\bibnamefont {Tanaka}}, \bibinfo {author}
  {\bibfnamefont {M.}~\bibnamefont {Fujiwara}}, \bibinfo {author}
  {\bibfnamefont {M.}~\bibnamefont {Sasaki}}, \bibinfo {author} {\bibfnamefont
  {A.}~\bibnamefont {Tajima}}, \bibinfo {author} {\bibfnamefont
  {S.}~\bibnamefont {Yorozu}}, \ and\ \bibinfo {author} {\bibfnamefont
  {A.}~\bibnamefont {Tomita}},\ }\href {\doibase 10.1364/OE.19.020531}
  {\bibfield  {journal} {\bibinfo  {journal} {Opt. Express}\ }\textbf {\bibinfo
  {volume} {19}},\ \bibinfo {pages} {20531} (\bibinfo {year}
  {2011})}\BibitemShut {NoStop}%
\bibitem [{\citenamefont {Yuan}\ \emph {et~al.}(2007)\citenamefont {Yuan},
  \citenamefont {Kardynal}, \citenamefont {Sharpe},\ and\ \citenamefont
  {Shields}}]{yuan:041114}%
  \BibitemOpen
  \bibfield  {author} {\bibinfo {author} {\bibfnamefont {Z.~L.}\ \bibnamefont
  {Yuan}}, \bibinfo {author} {\bibfnamefont {B.~E.}\ \bibnamefont {Kardynal}},
  \bibinfo {author} {\bibfnamefont {A.~W.}\ \bibnamefont {Sharpe}}, \ and\
  \bibinfo {author} {\bibfnamefont {A.~J.}\ \bibnamefont {Shields}},\ }\href
  {\doibase 10.1063/1.2760135} {\bibfield  {journal} {\bibinfo  {journal}
  {Applied Physics Letters}\ }\textbf {\bibinfo {volume} {91}},\ \bibinfo {eid}
  {041114} (\bibinfo {year} {2007})}\BibitemShut {NoStop}%
\bibitem [{\citenamefont {Yuan}\ \emph {et~al.}(2010)\citenamefont {Yuan},
  \citenamefont {Sharpe}, \citenamefont {Dynes}, \citenamefont {Dixon},\ and\
  \citenamefont {Shields}}]{yuan:071101}%
  \BibitemOpen
  \bibfield  {author} {\bibinfo {author} {\bibfnamefont {Z.~L.}\ \bibnamefont
  {Yuan}}, \bibinfo {author} {\bibfnamefont {A.~W.}\ \bibnamefont {Sharpe}},
  \bibinfo {author} {\bibfnamefont {J.~F.}\ \bibnamefont {Dynes}}, \bibinfo
  {author} {\bibfnamefont {A.~R.}\ \bibnamefont {Dixon}}, \ and\ \bibinfo
  {author} {\bibfnamefont {A.~J.}\ \bibnamefont {Shields}},\ }\href {\doibase
  10.1063/1.3309698} {\bibfield  {journal} {\bibinfo  {journal} {Applied
  Physics Letters}\ }\textbf {\bibinfo {volume} {96}},\ \bibinfo {eid} {071101}
  (\bibinfo {year} {2010})}\BibitemShut {NoStop}%
\bibitem [{\citenamefont {Restelli}\ and\ \citenamefont
  {Bienfang}(2012)}]{Restelli2012}%
  \BibitemOpen
  \bibfield  {author} {\bibinfo {author} {\bibfnamefont {A.}~\bibnamefont
  {Restelli}}\ and\ \bibinfo {author} {\bibfnamefont {J.~C.}\ \bibnamefont
  {Bienfang}},\ }\href@noop {} {\bibfield  {journal} {\bibinfo  {journal} {to
  be published in Proceedings of SPIE: Advanced Photon Counting Techniques VI}\
  }\textbf {\bibinfo {volume} {8375}} (\bibinfo {year} {2012})}\BibitemShut
  {NoStop}%
\bibitem [{\citenamefont
  {www.nano{-}tera.ch/nanoterawiki/QCRYPT}()}]{nanoTera2009}%
  \BibitemOpen
  \bibfield  {author} {\bibinfo {author} {\bibnamefont
  {www.nano{-}tera.ch/nanoterawiki/QCRYPT}},\ }\href@noop {} {}\BibitemShut
  {NoStop}%
\bibitem [{\citenamefont {Stucki}\ \emph {et~al.}(2005)\citenamefont {Stucki},
  \citenamefont {Brunner}, \citenamefont {Gisin}, \citenamefont {Scarani},\
  and\ \citenamefont {Zbinden}}]{stucki-2005-87}%
  \BibitemOpen
  \bibfield  {author} {\bibinfo {author} {\bibfnamefont {D.}~\bibnamefont
  {Stucki}}, \bibinfo {author} {\bibfnamefont {N.}~\bibnamefont {Brunner}},
  \bibinfo {author} {\bibfnamefont {N.}~\bibnamefont {Gisin}}, \bibinfo
  {author} {\bibfnamefont {V.}~\bibnamefont {Scarani}}, \ and\ \bibinfo
  {author} {\bibfnamefont {H.}~\bibnamefont {Zbinden}},\ }\href@noop {}
  {\bibfield  {journal} {\bibinfo  {journal} {Applied Physics Letters}\
  }\textbf {\bibinfo {volume} {87}},\ \bibinfo {pages} {194108} (\bibinfo
  {year} {2005})}\BibitemShut {NoStop}%
\bibitem [{\citenamefont {Stucki}\ \emph
  {et~al.}(2009{\natexlab{b}})\citenamefont {Stucki}, \citenamefont {Walenta},
  \citenamefont {Vannel}, \citenamefont {Thew}, \citenamefont {Gisin},
  \citenamefont {Zbinden}, \citenamefont {Gray}, \citenamefont {Towery},\ and\
  \citenamefont {Ten}}]{Stucki2009}%
  \BibitemOpen
  \bibfield  {author} {\bibinfo {author} {\bibfnamefont {D.}~\bibnamefont
  {Stucki}}, \bibinfo {author} {\bibfnamefont {N.}~\bibnamefont {Walenta}},
  \bibinfo {author} {\bibfnamefont {F.}~\bibnamefont {Vannel}}, \bibinfo
  {author} {\bibfnamefont {R.~T.}\ \bibnamefont {Thew}}, \bibinfo {author}
  {\bibfnamefont {N.}~\bibnamefont {Gisin}}, \bibinfo {author} {\bibfnamefont
  {H.}~\bibnamefont {Zbinden}}, \bibinfo {author} {\bibfnamefont
  {S.}~\bibnamefont {Gray}}, \bibinfo {author} {\bibfnamefont {C.~R.}\
  \bibnamefont {Towery}}, \ and\ \bibinfo {author} {\bibfnamefont
  {S.}~\bibnamefont {Ten}},\ }\href@noop {} {\bibfield  {journal} {\bibinfo
  {journal} {New Journal of Physics}\ }\textbf {\bibinfo {volume} {11}},\
  \bibinfo {pages} {075003} (\bibinfo {year} {2009}{\natexlab{b}})}\BibitemShut
  {NoStop}%
\bibitem [{\citenamefont {Branciard}, \citenamefont {Gisin},\ and\
  \citenamefont {Scarani}(2008)}]{Branciard2008}%
  \BibitemOpen
  \bibfield  {author} {\bibinfo {author} {\bibfnamefont {C.}~\bibnamefont
  {Branciard}}, \bibinfo {author} {\bibfnamefont {N.}~\bibnamefont {Gisin}}, \
  and\ \bibinfo {author} {\bibfnamefont {V.}~\bibnamefont {Scarani}},\ }\href
  {\doibase 10.1088/1367-2630/10/1/013031} {\bibfield  {journal} {\bibinfo
  {journal} {New Journal of Physics}\ }\textbf {\bibinfo {volume} {10}},\
  \bibinfo {pages} {013031} (\bibinfo {year} {2008})}\BibitemShut {NoStop}%
\bibitem [{\citenamefont {Hamkins}(2007)}]{Hamkins2007}%
  \BibitemOpen
  \bibfield  {author} {\bibinfo {author} {\bibfnamefont {J.}~\bibnamefont
  {Hamkins}},\ }\href {\doibase 10.1002/9781118256053.ch32} {\  (\bibinfo
  {year} {2007}),\ 10.1002/9781118256053.ch32}\BibitemShut {NoStop}%
\end{thebibliography}
%
%
\end{document}